\documentclass[12pt]{article}
\usepackage{amsmath}
\usepackage{amsfonts}   
\usepackage{amssymb}

\def\n{\noindent}

\begin{document}
\date{}
\title{{\bf{\Large A Note on the Thermodynamical Origin of Gravity}}}
\author{
{\bf {\normalsize Saurav Samanta}$
$\thanks{E-mail: srvsmnt@gmail.com}}\\
 {\normalsize Narasinha Dutt College,
129, Belilious Road, Howrah-711101, India}
\\[0.3cm]
}

\maketitle

\begin{abstract}
Following Mach's definition of mass and force in terms of space time measurement we build the conceptual structure of classical mechanics and show that it is remarkably similar to the results of Carnot and Clausius in thermodynamics. The analogy is extended to equality to derive the laws of gravity from the laws of thermodynamics. 
\end{abstract}

\section{Introduction}
If one tries to understand or predict the motion of a body, classical mechanics is of great help. This subject is often presented in an axiomatic manner where the basic assumptions are Newton's three laws of motion. These laws involve concepts like force, mass etc. which are not at all obvious. Even sometimes definitions of these terms are extracted from the laws itself. For example Newtons first law is interpreted as the definition of inertial frame and the second law is taken as the definition of force and mass simultaneously. This type of interpretations apparently involve logical circularity and hence their falsifiability is not easily understood\cite{raychaudhuri}. 

In his book Principia\cite{newton}, Newton did not consider his second law as the definition of force or mass. Instead, he defined these terms separately by abstract concepts. Since these definitions did not involve any operational methods, many scientists gave considerable efforts to put these concepts on solid philosophical ground\cite{jammer}. Among all these efforts, the approach taken by E. Mach\cite{mach} has a very significant effect on the further development of mechanics, most notably in the theory of relativity.

A striking feature of the general theory of relativity is the prediction of black holes from which nothing can escape. Further development of this subject shows that there is a close resemblance between black hole physics and thermodynamics\cite{bekenstein,hawking}. In fact it has been established that the entropy of a black hole is proportional to the area of the event horizon.  This is connected with the holographic principle where information of a certain amount of space resides on its boundary. 

These concepts have been developed further and basing on that, there are claims\cite{padmanabhan} that gravity may be an emergent phenomena instead of a fundamental force. In a recent paper\cite{verlinde} Newtonian law of gravity was obtained by interpreting it as an entropic force. In this formulation, force on a particle at some point was defied as the gradient of entropy times the temperature of that point. Einstein's theory of gravity was also interpreted in a similar manner. This concept has been elaborated in various works\cite{majhi,chang}.

In this article we show that the formulation of classical mechanics can be stated in a manner which is quite similar to the reasoning of thermodynamics\cite{zemansky}. We can interpret the results of Newtonian mechanics as theorems analogous to the theorems given by Carnot and Clausius. The study shows that mass of an object is similar to the inverse of temperature and force is similar to entropy. These similarity motivates us to interpret force as entropy (multiplied by appropriate constants) instead of entropy gradient. Subsequently we derive Newtonian gravity as well as Einstein's gravity from this point of view.

\n
\section{Laws of Mechanics}
The theory of classical mechanics gives us important clues about the general nature of motion in a simple form. Here by motion we mean the intuitive notion -- change of position with time. Before describing the laws let us start with few basic definitions.\\  
\underline{{\bf Definition 1: Particle}}\\
{\it Particle is an object of zero dimension.}\\
In classical mechanics, a body is supposed to be consists of many particles. So describing the motion of all individual particles is equivalent to describe the motion of the whole body.\\
\underline{{\bf Definition 2: Event}}\\
{\it Event is something which happens at a space time point.}\\
The motion of a particle is described by a series of events which are essentially measurement of the position of the particle with time. \\
\underline{{\bf Definition 3: Frame of Reference}}\\
{\it Frame of reference is a rigid coordinate system from which an event is described.}\\
Since the position of a particle depends on the choice of coordinates, it is necessary to fix the coordinate system from which the motion of a particle will be subsequently described.\\
\underline{{\bf Definition 4: Inertial Frame}}\\
{\it Inertial frame is a frame of reference in which at some time, a particle infinitely separated from other objects moves without any acceleration.}\\
As described by Mach\cite{jammer}, the laws of classical mechanics involves only kinematical properties of particles. This set of laws which gives us very powerful information about the acceleration of a particle will be stated for an inertial frame. In no way this is a limitation. In fact knowing these laws we can describe the motion of a body even in a non inertial frame.\\ 
\underline{{\bf Law 1: Law of No Self Interaction}}\\
Any particle infinitely separated from other objects, always moves without any acceleration, a particle obtains acceleration only in the presence of other objects.\\
\underline{{\bf Law 2: Law of Homogeneity and Isotropy of Space}}\\
If a stationary particle accelerates in presence of a second stationary particle, when both of them are isolated from other objects, it is always found that the second particle accelerates in the opposite direction. These accelerations of two particles act along the line joining two particles and their magnitude depend only on their relative separation.\\
\underline{{\bf Definition 5: Mass Function}}\\
{\it Mass Function of two particles is the negative reciprocal of the ratio of accelerations of these two particles, isolated from other objects, when the values of the accelerations are nonzero. }\\
If $a_A$ and $a_B$ are the acceleration of two bodies $A$ and $B$, when they are separated from other objects, then the mass function $M(A,B)$ is defined as
\begin{eqnarray}
M(A,B)=-\frac{a_B}{a_A}\label{0}
\end{eqnarray}
From the very definition, it follows that
\begin{eqnarray}
M(A,B)=\frac{1}{M(B,A)}\label{2}
\end{eqnarray}

\underline{{\bf Law 3: Law of Constancy of mass}}\\ 
Mass Function of two particles does not change with their motion and it is same as the ratio of mass functions of these two particles with a third particle.

 Using the previous notation for the mass function of two particles,  Law 2 can be mathematically stated as, 
\begin{eqnarray}
M(A,B)=\frac{M(A,C)}{M(B,C)} \label{1}
\end{eqnarray}
Using (\ref{2}), we write the above equation as
\begin{eqnarray}
M(A,B)=M(A,C)M(C,B) \label{3a}
\end{eqnarray}
We use the notation $m$ to denote the property of the body responsible for the above equation. Thus we get
\begin{eqnarray}
M(m_A,m_B)=M(m_A,m_C)M(m_C,m_B) \label{3}
\end{eqnarray}
Taking logarithm of the above equation and then double differentiating with respect to $m_A$ and $m_B$ we find
\begin{eqnarray}
\frac{\partial}{\partial m_A} \big[\frac{\partial}{\partial m_B}{\textrm {ln}}M(m_A,m_B)\big]=0\nonumber
\end{eqnarray}
The above equation naturally implies
\begin{eqnarray}
 \big[\frac{\partial}{\partial m_B}{\textrm {ln}}M(m_A,m_B)\big]=\phi(m_A)
\end{eqnarray}
Integrating the above differential equation and taking exponents, we get the most general solution as,
\begin{eqnarray}
M(m_A,m_B)=\psi(m_A)\chi(m_B)
\end{eqnarray}
This together with (\ref{3}) implies $\psi(m_C)\chi(m_C)=1$ for any $m_C$. Hence (\ref{0}) gives
\begin{eqnarray}
-\frac{a_B}{a_A}=\frac{\psi(m_A)}{\psi(m_B)}
\end{eqnarray}
Since the function $\psi(m)$ is independent of the kinematical behaviour, it can be used to define the mass $m$ in terms of the left hand side of the above equation. In that scale we define mass as,
\begin{eqnarray}
m=\psi(m)
\end{eqnarray}
and get
\begin{eqnarray}
-\frac{a_B}{a_A}=\frac{m_A}{m_B}\label{1000}
\end{eqnarray}
This equation which we name {\it mass theorem} defines the mass of a body with respect to any reference body mass of which can set as unity.\\
\underline{{\bf Law 4: Law of Instantaneous Interaction}}\\
Accelerations of two fixed particles, when isolated from other objects, depend only on the instantaneous kinematical properties of the particles and not on their past history.\\
\underline{{\bf Law 5: Law of Superposition of Accelerations}}\\
Isolated from other objects, when a system of particles interact among themselves, the acceleration of any individual particle is sum of the accelerations of that particle induced by other particles separately.\\
If we consider a system of interacting particles $A, B, C,\cdot\cdot\cdot$,separated from other bodies, then the acceleration of $A$ in presence of $B,C,\cdot\cdot\cdot$ is
\begin{eqnarray}
a_{A,BC\cdot\cdot\cdot}=a_{A,B}+a_{A,C}+\cdot\cdot\cdot\label{10}
\end{eqnarray}
where $a_{A,B}$ is the acceleration of $A$ due to $B$ only. We can get similar equations for $B,C\cdot\cdot\cdot$ as
\begin{eqnarray}
&&a_{B,AC\cdot\cdot\cdot}=a_{B,A}+a_{B,C}+\cdot\cdot\cdot\label{11}\\
&&a_{C,AB\cdot\cdot\cdot}=a_{C,A}+a_{C,B}+\cdot\cdot\cdot\label{12}\\
&&\cdot\nonumber\\
&&\cdot\nonumber\\
&&\cdot\nonumber
\end{eqnarray}
Multiplying the series of equations (\ref{10}), (\ref{11}), (\ref{12}), $\cdot\cdot\cdot$ by $m_A$, $m_B$, $m_C$, $\cdot\cdot\cdot$ and then adding we obtain
\begin{eqnarray}
m_Aa_{A,BC\cdot\cdot\cdot}+m_Ba_{B,AC\cdot\cdot\cdot}+m_Ca_{C,AB\cdot\cdot\cdot}+ \cdot\cdot\cdot=0\label{15}
\end{eqnarray}
where in the last step we have used the condition $m_Aa_{A,B}=-m_Ba_{B,A}$ for two interacting body $A$ and $B$. We define a quantity force by the relation
\begin{eqnarray}
F_{A}=m_Aa_{A,BC\cdot\cdot\cdot}\label{2000}
\end{eqnarray}
to write (\ref{15}) as
\begin{eqnarray}
F_A+F_B+F_C+\cdot\cdot\cdot=0
\end{eqnarray}
Since this equation has fundamental importance in further development of classical mechanics we give it a name -- {\it cancellation of internal forces}. Note that above equation of force is true only for a close isolated system. There is also one important equation of force which immediately follows from (\ref{10})
\begin{eqnarray}
F_{A,BC,\cdot\cdot\cdot}=F_{A,B}+F_{A,C}+\cdot\cdot\cdot
\end{eqnarray} 
We call this equation --{\it superposition of force}. Now we write (\ref{15}) in the following way
\begin{eqnarray}
-m_Aa_{A,BC\cdot\cdot\cdot}=m_Ba_{B,AC\cdot\cdot\cdot}+m_Ca_{C,AB\cdot\cdot\cdot}+ \cdot\cdot\cdot\label{20}
\end{eqnarray} 
Using the law of superposition of acceleration (\ref{11}), (\ref{12}), $\cdot\cdot\cdot$ in the right hand side of above equation we find
\begin{eqnarray}
-m_Aa_{A,BC\cdot\cdot\cdot}=m_Ba_{B,A}+m_Ca_{C,A}+ \cdot\cdot\cdot\label{21}
\end{eqnarray} 
Note that (\ref{20}) and (\ref{21}) are true true for any particle $A$, in presence of any arbitrary number of particles $B, C, \cdot\cdot\cdot$Now we consider a situation in which
\begin{eqnarray}
a_{B,AC\cdot\cdot\cdot}=a_{C,AB\cdot\cdot\cdot}= \cdot\cdot\cdot=a \ ({\textrm{say}})\label{30}
\end{eqnarray} 
In that case we write (\ref{20}) as
\begin{eqnarray}
-m_Aa_{A,BC\cdot\cdot\cdot}=(m_B+m_C\cdot\cdot\cdot)a\label{30a}
\end{eqnarray} 
We give further restriction on the system by taking their initial velocity ( $v^0$ ) to be same
\begin{eqnarray}
v_B^0=v_C^0=\cdot\cdot\cdot\label{31}
\end{eqnarray} 
and they are placed in a compact position
\begin{eqnarray}
r_B=r_C= \cdot\cdot\cdot\label{32}
\end{eqnarray} 
The last condition is the mathematical statement of rigid attachment from which (\ref{30}) and (\ref{31}) follow immediately. Under the above condition we can treat the system of particles $B,C,\cdot\cdot\cdot$ as a single particle and by comparing (\ref{30a}) and (\ref{1000}) we get the mass of the composite system $B,C,\cdot\cdot\cdot$ upon rigid attachment is
\begin{eqnarray}
M=m_B+m_C\cdot\cdot\cdot\label{33}
\end{eqnarray} 
We call the above result -- {\it additivity of mass}. Note that without the condition (\ref{32}), even if (\ref{30}) and (\ref{31}) are satisfied, we cannot get the total mass since the system under consideration is not a particle. Of course in that case and even in a more general case where any one (both) of (\ref{30},\ref{31}) is (are) violated, we can define the mass of the total system by the sum of masses of its individual components and can develop the concept of centre of mass. We do not discuss this point in the present article. 

Using (\ref{33}) we rewrite (\ref{30a}) as
\begin{eqnarray}
-m_Aa_{A,BC\cdot\cdot\cdot}=Ma\label{34}
\end{eqnarray} 
Now a comparison between (\ref{21}) and (\ref{34}) gives
\begin{eqnarray}
Ma=m_Ba_{B,A}+m_Ca_{C,A}+ \cdot\cdot\cdot
\end{eqnarray} 
Recalling the definition of force (\ref{2000}), we see, the left hand side of the above equation is the force ($\mathcal F$) on a system of particles attached rigidly and the right hand side is the sum of forces acting on individual particles. So the above result is written as, 
\begin{eqnarray}
\mathcal F=F_B+F_C+\cdot\cdot\cdot\label{40}
\end{eqnarray} 
 We name this theorem as -- {\it extensive nature of force} which is conceptually different from {\it superposition of force}{\footnote{Usually in textbook, this two theorems are not distinguished and one uses the parallelogram law of force which is taken as a fundamental principle\cite{synge}.}}. Note that there is nothing like extensive nature of acceleration, though there is a superposition principle for it. In this way, force, though initially defined in terms of the acceleration, turns out to be conceptually very much different from acceleration.

 The concept of  {\it extensive nature of force} gives us a deep insight about the general nature of properties of any body. In general, for a fixed environment, kinematical behavior of a body should depend on that body or in a different term, on the properties of that body. If we imagine a particle (with some definite property) to be consists of different particles (with different properties) so that sum of properties of individual particles will give the total property of the combined system, then (\ref{40}) will give severe restriction on the functional relationship between the properties of the body and the force acting on it. Let us now elaborate this point.
 
 We consider a set of unit mass particles $A, B, \cdot\cdot\cdot$ interacting with another particle $X$. Experience shows that, in general these particles $A, B, \cdot\cdot\cdot$ will have different accelerations even when their spatial separation with $X$ is same. This immediately suggests that the bodies under consideration $A, B, \cdot\cdot\cdot$ are not identical because a assumption about nature is that under a fixed condition, identical bodies must face identical results{\footnote{Though this assumption seems vary natural, it is wrong in the quantum domain\cite{feynman}.}}. So we can say one or more physical variables which are responsible for kinematical behavior have different values for different particles $A, B, \cdot\cdot\cdot$. We denote these variables by $q^i$. Thus the acceleration of any particle $A, B, \cdot\cdot\cdot$ can be written as,
\begin{eqnarray}
a=\xi(q^1,q^2,\cdot\cdot\cdot)
\end{eqnarray} 
Now we suppose there exists particles of unit mass for which only one of $q^i$ is nonzero. For those particles we can write the above equation as
\begin{eqnarray}
&&a_1=\xi(q^1,0,0,\cdot\cdot\cdot)\equiv\sigma^1(q^1) \ ({\textrm{say}})\label{60}\\
&&a_2=\xi(0,q^2,0,\cdot\cdot\cdot)\equiv\sigma^2(q^2) \ ({\textrm{say}})\label{61}\\
&&\cdot\label{62}\\
&&\cdot\\
&&\cdot
\end{eqnarray}
Now we consider two bodies $P'$ and $P''$ of type (\ref{60}). For them we can write (\ref{60}) as 
\begin{eqnarray}
&&a'_1=\sigma^1(q'^1)\label{61a}\\
&&a''_1=\sigma^1(q''^1)\label{62a}
\end{eqnarray} 
The ratio of above two equations gives
\begin{eqnarray}
\frac{a'^1}{a''^1}=\frac{\sigma^1(q'^1)}{\sigma^1(q''^1)}\label{70}
\end{eqnarray} 
Identifying
\begin{eqnarray}
\sigma^1(q^1)=q^1
\end{eqnarray} 
we write (\ref{70}) as
\begin{eqnarray}
\frac{a'^1}{a''^1}=\frac{q'^1}{q''^1}\label{71}
\end{eqnarray} 
If we consider a reference particle for which $q^1=1$ and $a^1=a^1_{{\textrm{ref}}}$ we can absolutely measure $q^1$ for any particle by the equation
\begin{eqnarray}
q^1= \frac{a^1}{a^1_{{\textrm{ref}}}}
\end{eqnarray} 
Now we consider $m^1$ number of identical particles of type (\ref{60}). The force acting on individual particle is given by,
\begin{eqnarray}
F=1.a^1=q^1{a^1_{{\textrm{ref}}}}
\end{eqnarray} 
Under rigid attachment the total force acting on this system is
\begin{eqnarray}
F^1_{{m}}=q^1_{{m}}{a^1_{{\textrm{ref}}}}
\end{eqnarray} 
Since we have proved mass is additive, the total mass of the combined system is $m$ and hence we can write the above equation as
\begin{eqnarray}
m^1a^1_m={q^1_m}a^1_{{\textrm{ref}}}\label{71a}
\end{eqnarray} 
where we denoted the acceleration of the system by $a^1_m$. Similar consideration for $m^2, m^3,\cdot\cdot\cdot$ number of particles of type (\ref{61},\ref{62},$\cdot\cdot\cdot$) will give
\begin{eqnarray}
&&m^2a^2_m={q^2_m}a^2_{{\textrm{ref}}}\label{72}\\
&&m^3a^3_m={q^3_m}a^3_{{\textrm{ref}}}\\
&&\cdot\\
&&\cdot\\
&&\cdot
\end{eqnarray}
Under rigid attachment of all these systems (\ref{71},\ref{72},$\cdot\cdot\cdot$) we have
\begin{eqnarray}
(\sum m_i)a=\sum q_m^ia^i_{{\textrm{ref}}}\label{73}
\end{eqnarray}
Now we can treat the above equation as the equation of motion of a single particle having mass $m$ and charges ($q^1_m,q^2_m,\cdot\cdot\cdot$). Thus (\ref{73}) reduces to
\begin{eqnarray}
ma=\sum q_m^ia^i_{{\textrm{ref}}}\label{74}
\end{eqnarray}
Above equation gives a remarkable information about the nature of force. If the quantities $q^1_m,q^2_m\cdot\cdot\cdot$ are kept constant, then the product of mass and acceleration is constant though neither of them has this property.\\\\

\section{Analogy Between the Concepts of Mechanics and Thermodynamics}
From the second law of thermodynamics one can derive Carnot's theorem which states that no engine can be more efficient{\footnote{If an engine takes $Q_1$ amount of heat, performs work $W$ and rejects $Q_2$ amount of heat, its efficiency is $\frac{W}{Q_1}$ or equivalently, $1-\frac{Q_2}{Q_1}$ since $Q_1=W+Q_2$}} than a reversible engine. This theorem can be used to define Kelvin temperature scale. If a reversible engine operates between reservoirs having temperature $T_1$ and $T_2$ ( $T_1>T_2$ ), then
\begin{eqnarray}
\frac{Q_1}{Q_2}=F(T_1,T_2)
\end{eqnarray}
Similarly if there is another reversible engine which operates between $T_2$ and $T_3$($T_2 >T_3$) for that engine
\begin{eqnarray}
\frac{Q_2}{Q_3}=F(T_2,T_3)
\end{eqnarray}
The second engine is taken in such a way that it absorbs the same amount of heat as rejected by the first engine. The net effect of this combined system is another engine for which
\begin{eqnarray}
\frac{Q_1}{Q_3}&=&\frac{Q_1}{Q_2}\frac{Q_2}{Q_3}F(T_1,T_2)\\
F(T_1,T_3)&=&F(T_1,T_2)F(T_2,T_3)\label{Th1}
\end{eqnarray}
This equation is analogous to (\ref{3}) which was used to define mass. In thermodynamics (\ref{Th1}) is used to derive the following relation for a reversible engine
\begin{eqnarray}
\frac{Q_1}{Q_2}=\frac{T_1}{T_2}\label{Th_2}
\end{eqnarray}
This result plays the same role as (\ref{1000}) i.e. we can define temperature from the above equation as we defined mass from (\ref{1000}). 

If we interpret $-Q_2$ as $Q_2$ amount of heat rejected in (\ref{Th_2}) then we get
\begin{eqnarray}
\frac{Q_1}{T_1}+\frac{Q_2}{T_2}=0
\end{eqnarray}
where $Q$ denotes the heat taken by the engine in different portions of its cycle. For an arbitrary cyclic process above equation can be generalized as
\begin{eqnarray}
\sum\frac{Q_i}{T_i}=0
\end{eqnarray}
Above equation is called Clausius theorem which is similar to the result (\ref{15}). From the Clausius theorem we see that the quantity $\frac{Q}{T}$ does not change for a cyclic reversible process and this quantity is defined as entropy ($S$) just as  $F=ma$ was defined in the first section.

One important difference between entropy and heat is that entropy, though defined in terms of heat, is a state function whereas heat is not. Using the extensive nature of force we have seen in section 2 that force and acceleration have the similar properties. We summarize the similarities between mechanics and thermodynamics in the following table(\ref{table1}).

\begin{table}[ht]
\begin{center}
\caption{Analogy between different concepts of classical mechanics and thermodynamics}

\label{table1}
~\\
\begin{tabular}{|c|c|c|}
\hline \hline
CLASSICAL THERMODYNAMICS & CLASSICAL MECHANICS \\ \hline 
1.Heat absorbed by the system ($Q$) & 1. Acceleration of a particle ($a$) \\ \hline
2. Carnot's theorem: $\frac{Q_1}{Q_2}=\frac{T_1}{T_2} $ & 2. {\it mass theorem}:$\frac{a_1}{a_2}=\frac{1/m_1}{1/m_2}$ \\ \hline
3. Claussius's result: $\sum\frac{Q_i}{T_i}=0$ & 3. {\it cancellation of internal forces}: $\sum m_ia_i=0$ \\ \hline
3. Entropy: $S=\frac{Q}{T}$ & 3. Force $F=ma$ \\ \hline 
4. For a composite system, $S=\sum S_i$ & 4. {\it extensive nature of force}: ${\cal{F}}=\sum F_i$\\ \hline
5. Entropy (not heat) is a state function& 5. Force (not acceleration) does not change for fixed charges.\\ \hline \hline
\end{tabular}
\end{center}
\end{table}

\pagebreak
\section{ Gravity as an Emergent Phenomena}
Since all particles follow the same trajectory in a gravitational field one cannot distinguish gravity and noninertial frame. This remarkable result is called the equivalence principle which is a consequence of the fact that, in the case of gravity charge is the inertial mass itself. Thus gravity is a very special type of force for which both force and acceleration are state functions in the sense that if we keep the mutual separations and charges of interacting particles fixed then there are no other parameter which can alter the force or acceleration of a particle. 

Now to find a thermodynamical mechanism behind gravity we must equate the various analogous quantities of mechanics and thermodynamics. Before doing that, we follow the basic motivation of \cite{verlinde} and suppose that the information of the energy within a spatial volume is stored on its boundary which can have finite degrees of freedom.

We assume that the entropy associated with each bit of the holographic screen is
\begin{eqnarray}
\Delta S=\frac{1}{2} k_B
\end{eqnarray}
so that it is consistent with the equipartition theorem which states that energy associated with each degree of freedom is $\frac{1}{2}k_BT$. Now according to our assumption the information of a certain volume is stored on the bits of its boundary. So if $n_i$ number of bits on the boundary have the energy $E_{n_i}$, then sum of $n_iE_{n_i}$ must be the total energy within the surface.
\begin{eqnarray}
\sum_i n_i E_{n_i}=E_{{\textrm{in}}}
\end{eqnarray}
When $n$ is large we can replace the discrete sum by integration
\begin{eqnarray}
\oint dn E(n)=E_{{\textrm{in}}}\label{main}
\end{eqnarray}
where the integration is performed over the complete surface. the above equation which is a direct consequence of the holographic principle has fundamental importance in the rest of the analysis. In order to use this equation we need to know the functional relationship between number of bits ($n$) and the surface area ($A$). We assume that $n$ is proportional to $A$
\begin{eqnarray}
dn=\frac{c^3}{G\hbar}dA\label{dn}
\end{eqnarray}
The proportionality constant is the inverse of the square of the Planck length ($\frac{1}{l_P^2}$). The above equation simply tells that each bit on the holographic screen has the area $l_P^2$. Now to derive Newton's law of gravity we recall that a freely falling observer finds the Unruh temperature
\begin{eqnarray}
T=\frac{1}{2\pi k_B}\frac{\hbar a}{c}
\end{eqnarray}
where $a$ is the acceleration of the observer. Thus the energy associated with a bit is
\begin{eqnarray}
E_n=T\Delta S=\frac{\hbar}{4\pi c}a\label{en}
\end{eqnarray}
Substituting (\ref{en}) and (\ref{dn}) in the left hand side of (\ref{main}) we find
\begin{eqnarray}
\frac{ac^2}{4\pi G}\oint dA=E_{{\textrm{in}}}
\end{eqnarray}
Now we consider a particle of mass $M$ is situated at the centre of a spherical holographic screen of radius $r$, then using the mass energy relation $E_{{\textrm{in}}}=Mc^2$ in the right hand side of the above equation we get
\begin{eqnarray}
\frac{a}{4\pi G}\oint dA=M
\end{eqnarray}
performing the surface integration we obtain the Newton's law of gravity
\begin{eqnarray}
a=\frac{GM}{r^2}
\end{eqnarray}
In order to derive Poisson's equation for arbitrary matter distribution we replace (\ref{en}) by the relation
\begin{eqnarray}
E_n=T\Delta S=\frac{\hbar}{4\pi c}\nabla\Phi\cdot\hat n\label{en1}
\end{eqnarray}
where $\Phi$ must be identified with the Newton's gravitational potential. In the above equation $\hat n$ is the unit vector normal to the holographic screen. Substituting (\ref{en1}) and (\ref{dn}) in the left hand side of (\ref{main}) we get
\begin{eqnarray}
\frac{c^2}{4\pi G}\oint\nabla\Phi\cdot\hat ndA=E_{{\textrm{in}}}
\end{eqnarray}
The right hand side of the above equation is the total energy contained within the surface. So it can be written as the volume integral of the mass density $\rho$. Thus,
 \begin{eqnarray}
\frac{c^2}{4\pi G}\oint\nabla\Phi\cdot\hat ndA=c^2\int\rho dv
\end{eqnarray}
Using the divergence theorem, above equation is written as
\begin{eqnarray}
\nabla^2\Phi=4\pi G\rho
\end{eqnarray}
which is the Poisson equation. This shows that if we identify the temperature of a bit on the screen properly, (\ref{main}) gives the correct nonrelativistic equation of gravity for arbitrary matter distribution.

The analysis presented above can also be generalized for the relativistic case. For that we consider a global time like killing vector $\xi^a$ in a static background. This is related to the generalized gravitational potential in the following way,
\begin{eqnarray}
\Phi=\frac{1}{2}{\textrm{ln}}\left(-\xi^a\xi_a\right)
\end{eqnarray}
The 4-velocity $u^a$ can be written in terms of $\xi^a$ as
\begin{eqnarray}
u^b=e^{\Phi}\xi^b
\end{eqnarray}
The above equation gives the expression for 4-acceleration $a^b$ as,
\begin{eqnarray}
a^b\equiv u^a\nabla_au^b=e^{-2\Phi}\xi^a\nabla_a\xi^b=\nabla^{b}\Phi
\end{eqnarray}
The generalized Unruh temperature on the screen is given by,
\begin{eqnarray}
T=\frac{\hbar a}{2\pi k_B c}=\frac{\hbar N^b\nabla_b\Phi}{2\pi k_Bc}
\end{eqnarray}
Since the temperature is measured with respect to the infinitely separated reference point, we put a prefactor $e^{\Phi}$ (redshift factor) in the numerator of the above expression to get
\begin{eqnarray}
T=\frac{\hbar e^{\Phi} N^b\nabla_b\Phi}{2\pi k_Bc}
\end{eqnarray}
This gives the energy of a bit
\begin{eqnarray}
E_n=T\Delta S=\frac{\hbar e^{\Phi} N^b\nabla_b\Phi}{4\pi c}
\end{eqnarray}
Now substituting the above expression in the left hand side of (\ref{main}) and then using (\ref{dn}) we obtain,
\begin{eqnarray}
\frac{c^2}{4\pi G}\oint e^{\Phi}N^b\nabla_b\Phi dA=E
\end{eqnarray}
The left hand side of the above equation can be written in terms of the killing vector $\xi^a$ and the right hand side can be identified with the expression of Komar mass to get
\begin{eqnarray}
\frac{1}{8\pi G}\oint dx^a\wedge dx^b\epsilon_{abcd}\nabla^c\xi^d=2\int\left(T_{ab}-\frac{1}{2}Tg_{ab}\right)n^a\xi^bdv
\end{eqnarray}
where $T_{ab}$ is the energy momentum tensor and $n^a$ is the unit vector normal to the holographic screen. Now making use of the Stokes theorem in the surface integration term we find,
\begin{eqnarray}
\frac{1}{4\pi G}\int R_{ab}n^a\xi^bdv=2\int\left(T_{ab}-\frac{1}{2}Tg_{ab}\right)\label{einstein}
\end{eqnarray} 
where we have used the relation 
\begin{eqnarray}
\nabla^a\nabla_a\xi^b=-R^b_{ \ a}\xi^a
\end{eqnarray}
Since (\ref{einstein}) is true for any arbitrary volume element, we obtain the Einstein equation
\begin{eqnarray}
R_{ab}=8\pi G\left(T_{ab}-\frac{1}{2} Tg_{ab}\right)
\end{eqnarray}
Thus we see that, Newton's law of gravity as well as Einstein's gravity can be derived from the holographic principle even without interpreting force as the entropy gradient.
\section{ Conclusions}
The laws of mechanics in an inertial frame have been stated from the point of view of Mach. These laws were then used to define mass of an object and force. We then derived some important results of classical mechanics like additivity of mass and  extensive nature of force. The concept of force led to the understanding of charge. The structure of the subject is similar to that of thermodynamics. Various concepts of mechanics and thermodynamics are found to be analogous. Instead of defining force as entropy gradient we defined it in terms of entropy and using the holographic principle we recovered both the Newton's law of gravity and Einstein's general relativity. Thus the close resemblance between mechanics and thermodynamics, discussed in this paper, indicates that gravity may not be a fundamental force and there may be a statistical description behind it. 
\vskip 1.5cm

{\bf Acknowledgment}\\

We thank Dr. T. Sinha for useful discussions.

\end{document}